\documentclass{jps-cp}
\usepackage{txfonts} 
\usepackage{wrapfig}
 
\title{Magneto-Optics of the Weyl Semimetal TaAs in the THz and IR Regions}

\author{
Shin-ichi Kimura$^{1,2}$, Yuko Yokoyama$^{2}$, Yuki Nakajima$^{2}$, Hiroshi Watanabe$^{1,2}$,  \\
J\"org Sichelschmidt$^{3}$, Vicky S\"u\ss$^{3}$, Marcus Schmidt$^{3}$, Claudia Felser$^{3}$
}

\inst{
$^{1}$Graduate School of Frontier Biosciences, Osaka University, Suita 565-0871, Japan\\
$^{2}$Department of Physics, Osaka University, Toyonaka 560-0043, Japan\\
$^{3}$Max Planck Institute for Chemical Physics of Solids, 01187 Dresden, Germany
}

\email{kimura@fbs.osaka-u.ac.jp}

\recdate{\today}

\abst{
The magnetic-field dependence of optical reflectivity [$R(\omega)$] and optical conductivity [$\sigma(\omega)$] spectra of the ideal type-I Weyl semimetal TaAs has been investigated at the temperature of 10~K in the terahertz (THz) and infrared (IR) regions.
The obtained $\sigma(\omega)$ spectrum in the THz region of $\hbar\omega\leq15$~meV is strongly affected by the applied magnetic field ($B$): 
The Drude spectral weight is rapidly suppressed and an energy gap originating from the optical transition in the lowest Landau levels appears with a gap size that increases in proportion to $\sqrt{B}$, which suggests linear band dispersions.
The obtained THz $\sigma(\omega)$ spectra could be scaled not only in the energy scale by $\sqrt{B}$ but also in the intensity by $1/\sqrt{B}$ as predicted theoretically.
In the IR region for $\hbar\omega\geq17$~meV, on the other hand, the observed $R(\omega)$ peaks originating from the optical transitions in higher Landau levels are proportional to linear-$B$ suggesting parabolic bands.
The different band dispersions originate from the crossover from the Dirac to the free-electron bands.
}

\kword{Weyl semimetal, TaAs, optical conductivity, magneto-optics, terahertz, infrared}

\begin{document}
\maketitle

\section{Introduction}

Weyl fermions as well as Weyl semimetals are massless chiral fermions appearing in solids and are attracting attention not only in the relation to elementary particles, especially neutrinos, but also having characteristic physical properties such as a giant magneto-resistance, chiral anomaly, and the appearance of Fermi arcs on the surfaces~\cite{Bernevig2015, Jia2016}.
The properties originate from a pair of Weyl points (two Dirac points) without spin degeneracy owing to the space/time reversal symmetry breaking that exists near the Fermi level ($E_{\rm F}$).
To clarify the electronic structure including the energies of Weyl points and the saddle points between the Weyl point pair, 
bulk-sensitive angle-resolved photoelectron spectroscopy~\cite{Lv2015,NXu2016}, optical conductivity [$\sigma(\omega)$]~\cite{BXu2016,Neubauer2018}, NMR~\cite{Yasuoka2017} and others have been performed so far.
We have also reported the temperature dependence of the optical conductivity [$\sigma(\omega)$] spectrum of pronounced type-I Weyl semimetals Ta$Pn$ ($Pn$~=~As, P)~\cite{Kimura2017}.
In comparison with other types of Weyl semimetals, for instance the type-II Weyl semimetals WTe$_2$ and MoTe$_2$~\cite{Kimura2019}, type-I Weyl semimetals have a simple electronic structure because the Fermi surfaces mainly consist of Dirac bands.
The Weyl points of TaAs are located very close to $E_{\rm F}$ and the saddle points between a Weyl point pair are consistent with band calculations based on a local-density approximation (LDA).
Therefore, we have concluded that TaAs is a candidate for typical type-I Weyl semimetals.

Related to the giant magneto-resistance of Weyl semimetals, the electronic structure can be easily modulated by magnetic fields.
Theoretical expectations for the appearance of Landau levels in the Weyl bands near $E_{\rm F}$ have been reported by Ashby and Carbotte~\cite{Ashby2013}.
According to the theory, both peak energy and peak intensity can be scaled by the magnetic field.

In this paper, we report the magnetic-field dependence of the electronic structure of an ideal Weyl semimetal TaAs by using $\sigma(\omega)$ spectra in the terahertz (THz) and infrared (IR) regions at the temperature of 10~K.
As a result, the obtained THz $\sigma(\omega)$ spectra show an energy gaps with the gap sizes being proportional to $\sqrt{B}$, suggesting the linear band dispersion.
The peak intensity of the gap structure is proportional to $1/\sqrt{B}$.
This allows scaling of spectral shapes and intensities in magnetic fields confirming the theoretical expectations.
In the IR region, some peaks are related to optical absorptions from higher-energy Landau levels
with energies varying linear in B. 
This observation suggests a change in the band dispersion from a Dirac-like linear to a free-electron-like parabolic one from near-$E_{\rm F}$ to higher energy states.

\section{Experimental}
Single-crystalline TaAs samples were synthesized by the chemical vapor growth method.
Near-normal incident reflectivity [$R(\omega)$] spectra in the $ab$ plane were acquired over a wide photon-energy $\hbar\omega$ range of 3~meV--30~eV to ensure accurate Kramers-Kronig analysis (KKA)~\cite{Kimura2013}.
$R(\omega)$ spectra in the THz and IR regions under magnetic fields were acquired in the configuration of $\vec{E} \perp \vec{B} \parallel c$ axis, 
where $\vec{E}$ and $\vec{B}$ are the electric vector of incident light and the applied magnetic field vector, by using Martin-Puplett-type and Michelson-type rapid-scan Fourier spectrometers, respectively, and a superconducting magnet for photon energy ranges of 3~meV~$\leq\hbar\omega\leq$~25~meV and 10~meV~$\leq\hbar\omega\leq$~54~meV at the temperature of 10~K~\cite{Kimura2002}.
$\sigma(\omega)$ spectra were obtained from $R(\omega)$ via KKA using appropriate extrapolations (the Hagen-Rubens function for $B\leq2$~T and a constant value for $B\geq4$~T) of the obtained spectra.

\section{Analysis and Discussion}

\begin{figure}[t]
 \begin{center}
  \includegraphics[width=70mm]{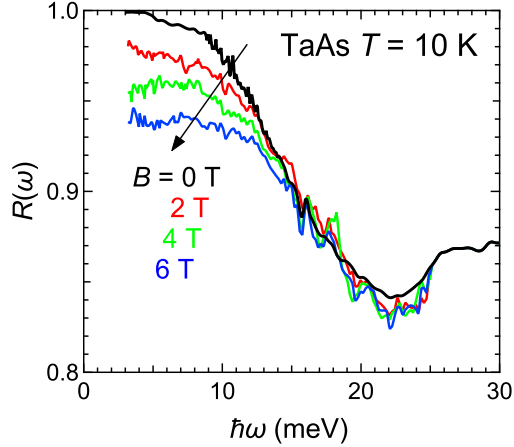}
\caption{
Magnetic-field ($B$) dependence of reflectivity [$R(\omega)$] spectrum of TaAs in the energy range of $\hbar\omega\leq25$~meV at the temperature of 10~K.
$R(\omega)$ spectrum for $\hbar\omega\geq25$~meV was taken only at $B=0$~T.
}
 \end{center}
\label{fig1}
\end{figure}

The obtained magnetic-field-dependent $R(\omega,B)$ spectra in the THz region are plotted in Fig.~1.
Temperature-dependent $R(\omega)$ spectra reported previously~\cite{Kimura2017} suggest that a large Drude weight appearing at high temperatures changes to a Drude component with low carrier density and long relaxation time with decreasing temperature.
The residual Drude component originates from carriers of Weyl points.
The $R(\omega,B={\rm 0~T})$ intensity at the accessible lowest photon energy of 3~meV is almost unity, which corresponds to the good metallic character with a direct current conductivity ($\sigma_{DC}$) as high as $10^6~\Omega^{-1}{\rm cm}^{-1}$~\cite{Zhang2015}.
The $\sigma_{DC}$ is suppressed with increasing magnetic field.
At the same time, the $R(\omega)$ value at $\hbar\omega\sim5$~meV decreases with applying magnetic field suggesting the decrease of $\sigma(\omega)$.

\begin{figure}[t]
\begin{center}
\includegraphics[width=1\textwidth]{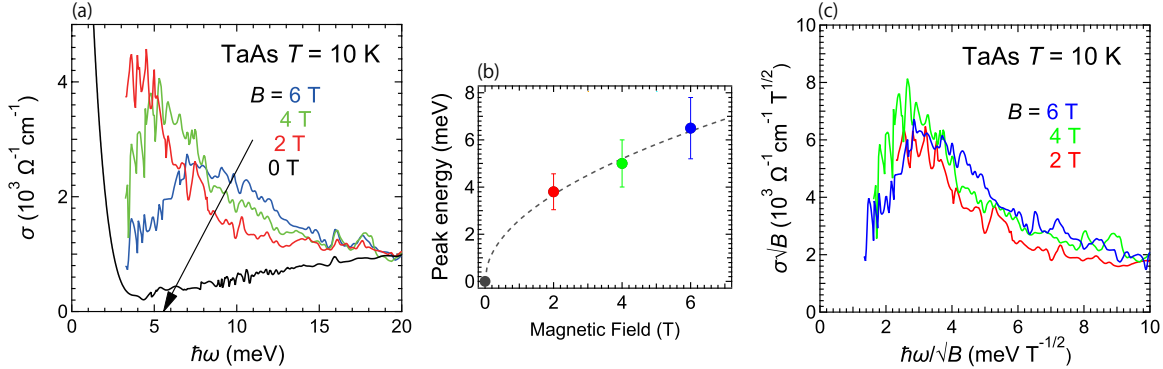}
\caption{
(a) Magnetic-field dependence of optical conductivity [$\sigma(\omega,B)$] spectrum of TaAs in the THz region at the temperature of 10~K.
The spectrm of $B=0$~T below 3~meV is in the extrapolation region with the Ha
(b) Energy of the broad peak in (a) as a function of magnetic field.
The dashed line is the fitted function of $\hbar\omega_{peak}=a\sqrt{B}$, where $a=2.65$ [meV/T$^{1/2}$].
(c) Same as (a), but the horizontal and vertical axes are divided by $\sqrt{B}$ and $1/\sqrt{B}$, respectively.}
\end{center}
\label{fig2}
\end{figure}

$\sigma(\omega,B)$ spectra derived from KKA of $R(\omega,B)$ spectra are shown in Fig.~2a.
At $B=0$~T,  the behavior $\sigma(\omega)\propto\omega$ appearing in $\hbar\omega\geq5$~meV suggests optical transitions in Dirac bands~\cite{Hosur2012}.
There is a rapid decrease of the Drude tail for $\hbar\omega\leq5$~meV, which is consistent with the $\sigma_{DC}$.
The Drude component is strongly suppressed with increasing magnetic field corresponding to the giant magneto-resistance~\cite{Huang2015}, and in addition, the intensity of the tail at $\hbar\omega\sim10$~meV increases.
At $B=2$~T, the $\sigma(\omega)$ for $\hbar\omega\leq5$~meV becomes almost constant.
The $\sigma_{DC}$s at $B\geq1$~T and $T=10$~K is lower than $1\times10^{3}~\Omega^{-1}{\rm cm}^{-1}$~\cite{Zhang2015}, 
the $\sigma(\omega,$B=2$~T)$ spectrum seems to have a peak at $\sim4$~meV.
As magnetic field increases, a broad peak becomes clearly visible.
The peak position shifts to the high energy side from $\sim5$~meV at 4~T to $\sim7$~meV at 6~T.
The peak energies plotted as a function of magnetic field with a reproducing guide line of $\hbar\omega_{peak}\propto\sqrt{B}$ are shown in Fig.~2b.
This suggests that the peak originates from the optical transition in Landau levels in a linear band dispersion of Weyl bands, which is consistent with the $\sigma(\omega)\propto\omega$ behavior.
Similar behaviors have been observed in Dirac semimetals of ZrTe$_5$ and Cd$_3$As$_2$~\cite{Chen2015,Hakl2018} and also in a Weyl semimetal NbAs~\cite{Yuan2018}.
However, the scaling in the intensity of the $\sigma(\omega)$ peak is firstly reported experimentally, to our best knowledge. 

Thanks to the theory by Ashby and Carbotte~\cite{Ashby2013}, for optical absorption peaks from the Landau levels of Weyl bands, 
not only the peak energies of the $\sigma(\omega,B)$ spectra at different magnetic field are scaled by $\sqrt{B}$ but also the peak intensity is scaled by $1/\sqrt{B}$.
In addition, the peak shape is expected to be asymmetric showing a rapid increase of the lower energy side and a tender slope at the higher energy side of the peak.
The obtained $\sigma(\omega,B)$ spectra shown in Fig.~2a are replotted in Fig.~2c with horizontal and vertical axes divided by $\sqrt{B}$ and $1/\sqrt{B}$, respectively.
All of the $\sigma(\omega,B)$ spectra seem to be similar to one another, which suggests that not only the peak energy and intensity but also the peak shape are consistent with the theoretical expectations.
Therefore, the observed peaks in the $\sigma(\omega,B)$ spectra are concluded to originate from the absorption peaks from the Landau levels of Weyl bands. 

\begin{figure}[t]
\begin{center}
\includegraphics[width=0.8\textwidth]{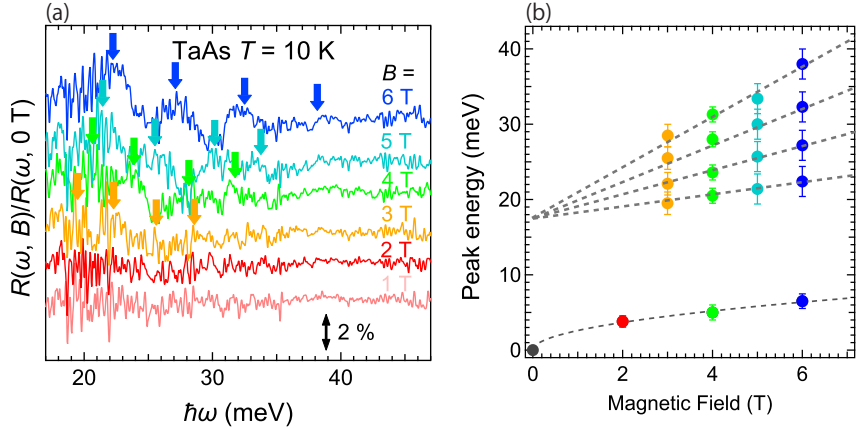}
\caption{
(a) $R(\omega)$ spectra of TaAs under magnetic field divided by the spectrum at 0~T.
The spectra are shifted by 2~\% upwards by equal intervals.
(b) Peak energies observed in Figs.~2a and 3a as a function of magnetic field.
The dashed lines are guides for the eyes.
}
\end{center}
\label{fig3}
\end{figure}
Absorptions from the Landau levels are also observed in the IR region.
Since the spectral changes are smaller than $\sim2$~\% in this energy region, the relative $R(\omega,B)$ spectra at higher magnetic fields divided by the 0~T data are used to obtain the magnetic-field dependence as shown in Fig.~3a.

Down arrows in Fig.~3a are identified peaks corresponding to optical absorptions from the Landau levels.
The peak energies are plotted as a function of magnetic field shown in Fig.~3b, where the peaks in the THz region (same as Fig.~2b) are also plotted.
A $B$-linear dependence of the peak energy in the IR region is clearly observed, which is different from the $\sqrt{B}$-dependence of the THz peak.
The $B$-linear dependence suggests a free-electron-like parabolic band dispersion ($\omega \propto k^2$) appearing in the IR energy range.

When these peaks are extrapolated to 0~T, they converged to $\sim17$~meV, 
which is close to the energy difference between saddle points of Weyl bands of $W_2$ points~\cite{Kimura2017}.
Near the Weyl points, the band dispersion is linear because of the Dirac band shape, but far from the Weyl points (near the saddle points), the band dispersions become complex, but similar to free-electron bands.
Therefore, the observation of the $\sqrt{B}$- and $B$-dependences of the absorption peaks from the Landau levels in the THz and IR regions, respectively, suggests a transformation from the Dirac linear dispersions to the free-electron-like parabolic band dispersions at about 17~meV.

\section{Conclusion}

Reflectivity as well as optical conductivity [$\sigma(\omega,B)$] spectra of the typical type-I Weyl semimetal TaAs have been measured under magnetic field at the temperature of 10~K to investigate the shape of the conduction bands using the magnetic-field dependence of the absorption peaks from Landau levels.
The Drude peak was strongly suppressed with increasing magnetic field, which corresponds to the giant magneto-resistance.
A $\sqrt{B}$ dependence of the peak energy in the THz region below 10~meV was observed and the $\sigma(\omega,B)$ spectral intensity at each magnetic field was scaled by $1/\sqrt{B}$ as proposed by theory.
On the other hand, in the IR region above about 17~meV, a linear-$B$ dependence of the peak energy appears suggesting a parabolic free-electron-like band dispersion.
The observation suggests a transformation from a Dirac linear band dispersion to a free-electron-like parabolic dispersion at about 17~meV.

\section*{Acknowledgments}

We would like to acknowledge UVSOR staff members for synchrotron radiation experiments.
Part of this paper was supported by the Use-of-UVSOR Facility Program (BL7B, 2015) of the Institute for Molecular Science. 



\end{document}